\documentclass[aps,twocolumn,groupedaddress,superscriptaddress,floatfix,pra]{revtex4-1}
\usepackage{graphicx}
\usepackage{dcolumn}
\usepackage{bm}
\usepackage{amsmath}
\usepackage{amsfonts}
\usepackage{amssymb}
\usepackage{dsfont}
\usepackage{color}
\usepackage{xcolor}
\usepackage{siunitx}
\usepackage[colorlinks=true,
    linkcolor=blue,
    filecolor=blue,      
    urlcolor=blue,
    citecolor = blue]{hyperref}
\usepackage{braket}
\usepackage{mathtools}
\usepackage[normalem]{ulem}
\usepackage{scalerel}
\usepackage{ bbold }
\usepackage{inputenc}
\usepackage[T1]{fontenc}

\newcommand{\ag}[1]{{\color{blue} #1 }}

\begin{document}

\title{Storage properties of a quantum perceptron} 

\author{Aikaterini (Katerina) Gratsea}
\affiliation{ICFO - Institut de Ci\`{e}ncies Fot\`{o}niques, The Barcelona Institute of Science and Technology, Av. Carl Friedrich Gauss 3, 08860 Castelldefels (Barcelona), Spain}
\author{Valentin Kasper}
\affiliation{ICFO - Institut de Ci\`{e}ncies Fot\`{o}niques, The Barcelona Institute of Science and Technology, Av. Carl Friedrich Gauss 3, 08860 Castelldefels (Barcelona), Spain}
\author{Maciej Lewenstein}
\affiliation{ICFO - Institut de Ci\`{e}ncies Fot\`{o}niques, The Barcelona Institute of Science and Technology, Av. Carl Friedrich Gauss 3, 08860 Castelldefels (Barcelona), Spain}
\affiliation{ICREA, Pg. Llu\'{\i}s Companys 23, 08010 Barcelona, Spain}

\date{\today}
\begin{abstract} 
Driven by growing computational power and algorithmic developments, machine learning methods have become valuable tools for analyzing vast amounts of data. Simultaneously, the fast technological progress of quantum information processing suggests employing quantum hardware for machine learning purposes. Recent works discuss different architectures of quantum perceptrons, but the abilities of such quantum devices remain debated. Here, we investigate the storage capacity of a particular quantum perceptron architecture by using statistical mechanics techniques and connect our analysis to the theory of classical spin glasses. Specifically, we focus on one concrete quantum perceptron model and explore its storage properties in the limit of a large number of inputs.
\end{abstract}

\maketitle

\section{Introduction \label{Introduction}}

\noindent{\it Importance of machine learning.} The rapid development of machine learning algorithms revolutionized our day-to-day lives and created novel connections between such diverse fields as computer science and neuroscience~\cite{neuroscience}, physics~\cite{CarleoRMP}, and engineering~\cite{engineering}. At the core of the success of machine learning  are deep artificial neural networks~\cite{deep_learning}. Whereas artificial neural networks initially had a biological motivation~\cite{rosenblatt1958perceptron}, the modern perspective considers artificial neural networks as a form of information processing~\cite{minsky69perceptrons, astroML, associative_memory, classification}. \\


\noindent{\it The quest for quantum advantage in quantum learning.} Machine learning algorithms have to be very efficient to handle a vast amount of information. This quest for efficiency created considerable interest to implement neural networks in dedicated hardware~\cite{neuromorphic}. However, the usage of quantum hardware, especially quantum computers, motivates  the question: How to use quantum mechanics best for machine learning purposes~\cite{Biamonte2017, Schuld2015, Schuld2014,  Muller1995, Shcherbina2001RigorousSO, TS2}? The quest for efficiency eventually led to the field of quantum machine learning, which now encompasses even more aspects \cite{Briegel}, e.g., the application of machine learning techniques to analyze quantum systems and devices (cf. \cite{Carleo, CarleoRMP, bookWittek, lewenstein-phase}), but also the direct implementation of machine learning concepts on quantum hardware itself~\cite{Schuld2014, huang2021power}. In this work, we will be concerned with the latter perspective. One should fairly admit, however, that in the absence of large scale quantum computers with fault tolerant error correction, in the NIQS era~\cite{Preskill18}, the question "Is quantum advantage the right goal for quantum machine learning?" is open~\cite{Schuld-failure, synergy}.\\


\noindent{\it Statistical mechanics of machine learning.} Statistical mechanics was introduced to study classical neural networks  (NN)in the 1980s, mostly focusing on Hopfield-like attractor ANN~\cite{Hopfield,Amit}.  Then, Hopfield networks were studied as constraint satisfaction networks and also cognitive models, e.g. as a simple model for memory, but have never obtained any relevance for machine learning. Their relationship with contemporary machine learning is typically considered to be remote. Restricted Boltzmann machines are related to them and do not fall out of fashion completely (cf. \cite{Decelle,Pozas} and references therein), but their working only partially relies on capacity estimates for Hopfield networks. 

Recently, this situation starts to change with the progress of statistical methods that can be applied to classical feed-forward NNs and deep learning models \cite{Tishby}. Initially, feed forward NNs and deep learning (cf. \cite{RumelhartMcClellandGroup86, McClellandRumelhartGroup86}) were considered very separately from ANNs, but nowadays statistical physics methods are being applied to feed forward NNs (\cite{Tishby,CarleoRMP}. ANNs, and even the simple perceptrons, are back in the centre of interests  with the advent of quantum technologies, due to the possible realization of simple quantum NNs with ultra-cold atoms, trapped ions, Rydberg atoms, super-conducting qubits, or photonic systems  (cf. \cite{lnp1000}). Importantly, tools of statistical physics have already established deep relations between neural networks, spin glasses, complexity, and information processing~\cite{Nishimori2001, Muller1995}. One advantage of statistical physics is the computation of global properties of physical systems without knowing the microscopic details. Recently, there has been a true revival of  increasing interest in using statistical physics techniques to study quantum information problems~\cite{symmetricBinaryPerceptrons, CarleoRMP}.

One important application of statistical physics to information
processing concerns the Hopfield-like networks or even simpler perceptrons. For example, there are many learning rules~\cite{review:learning_rules} for the Hopfield network, which can be used to obtain a desired input-output relations. Still, the connection between the global properties of the perceptron, and a specific learning rule might be challenging to analyze.\\

\noindent{\it Gardner's program.} In the seminal works~\cite{Gardner1988a,GD} Gardner addressed this challenge by using statistical physics to calculate the maximum storage capacity of a Hopfield network without referring to any specific learning rule. This approach of analyzing artificial networks without specifying the learning rule and treating the weights as a random 
variable is frequently referred to as Gardner's program. 

So far, the maximum storage capacity has also been considered for other QNN models~\cite{Hebbrule, symmetricBinaryPerceptrons, ding2018capacity}. In this work we apply Gardner's program to a specific quantum perceptron architecture proposed in~\cite{Tacchino2019}, where the authors discuss a specific quantum analog of the classical perceptron - the building block of neural networks. This quantum perceptron model has a direct implementation on quantum hardware. We focus here on the maximal storage capacity of this proposal without referring to any learning rule.\\

\noindent{\it Gardner's program and quantum NNs.} The work of Gardner demonstrated that we can study Hopfield networks independent of the precise learning rule that is used. But, it is not only of  considerable historical significance: it sheds light on the most important questions of contemporary quantum ML \cite{Schuld-failure}. Moreover, it is extremely general, adaptive and versatile; so far it has been applied for very different models of quantum perceptrons or quantum neural networks, or even to calculate volumes of quantum correlated (entangled) states. Here is a list of selected examples and applications of Gardner's program for various models of quantum perceptrons, quantum NN and more:
\begin{itemize}
\item In ref.~\cite{QP_Lewenstein} quantum perceptron is defined as a unitary map followed by projective measurments in a multidimensional Hilbert space. Calculation of the relative volume reduces to calculation of the volume in the unitary group space. 
\item Gardner's relative volume approach clearly inspired the pioneering attempts to estimate volume of quantum correlated states, such as entangled states \cite{Zyczkowski}. Integration consists in the first place in integration over unitary group in very high dimension, though.
\item Recently, Gardner's program has been used on QNN models~\cite{lewenstein2020storage} corresponding to completely positive trace preserving maps (CPTP). Here relative volume requires integration over the space of maps.
\item Gardner's program inspired investigation of  the relative volume of parent Hamiltonians having a target ground state up to some fixed error $\epsilon$ \cite{Anna}.
\item A careful look on other models of quantum perceptrons immediately suggests that Gardner's approach is possible and might turn out to be useful as well. For instance, for  the quantum perceptron models introduced in~\cite{Erik_perceptron, Beer_2020} integration over the unitary group is  needed to realize the Gardner's program. 
\item In the present paper, Gardner's program is applied to the specific quantum perceptron model proposed in~\cite{Tacchino2019}.
\end{itemize}

By studying these models, we are one step closer to understand whether they provide a hope for quantum advantage. \\


\noindent{\it Plan of the paper.} This article is structured as follows: After motivating Gardner's program in this section, we discuss Gardner's program in detail and apply it to a quantum perceptron architecture in Sec.~\ref{sec:Results}. In Sec.~\ref{sec:Discussion}, we discuss the main result: the calculation of the storage capacity of a quantum perceptron by applying statistical physics techniques. Finally, we give the computational details of Gardner's program for the quantum perceptron in Sec.~\ref{methods}.

\section{Quantum Perceptrons and Gardner's program \label{sec:Results}}

\subsection{Classical perceptron model}
A classical perceptron is a function that maps a $N$ dimensional input $\vec{i}^{\mu} = (i^{\mu}_1, \ldots, i^{\mu}_{N})^T$ onto an output $\sigma^{\mu}$, where the weight vector $\Vec{w} =(w_1, \ldots, w_{N})^T$ determines the information
processing. The additional label $\mu \in \{1, 2, ..., p\}$ denotes different pairs of input vectors and outputs~\cite{Nishimori2001, minsky69perceptrons}.
Moreover, we consider the following activation function
\begin{equation} \label{clas_act}
    \sigma^{\mu}= \theta \left( \vec{i}^{\mu} \cdot \vec{w} - \kappa \right) \, ,
\end{equation} 
where $\kappa$ is the threshold and $\theta(\cdot)$ is the Heaviside function realizing the non-linearity of the perceptron model, see Fig.~\ref{Fig1}a.\\

\subsection{Quantum perceptron model}
\begin{figure}
\includegraphics[scale=1.0]{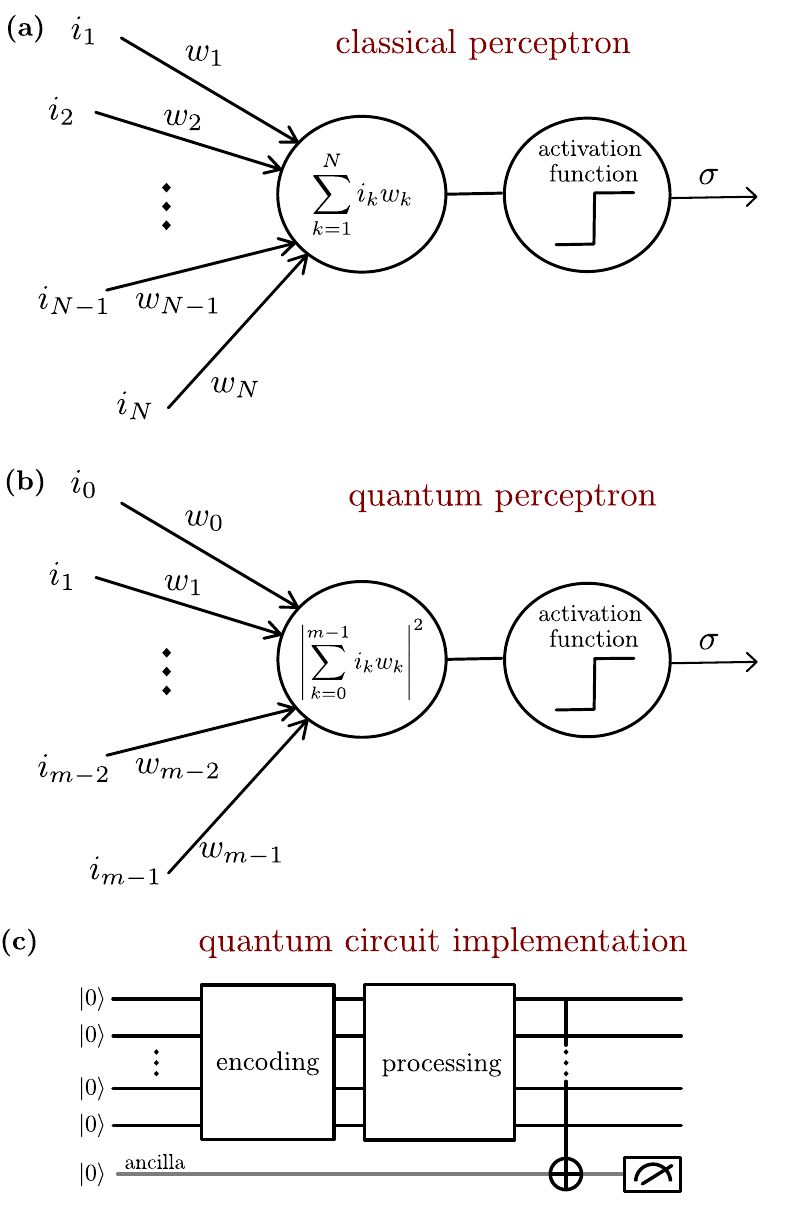}
\caption{\textbf{Classical and quantum perceptrons} 
\textbf{(a)} Schematic outline of the classical perceptron: An $N$-dimensional input array $\vec{i}$ is processed with a weight vector $\vec{w}$ such that $\vec{i}\cdot \vec{w}$ enters the activation function. 
\textbf{(b)} Schematic outline of the quantum perceptron: An $m$-dimensional input array $\vec{i}$ is processed with a weight vector $\vec{w}$ to produce the inner product squared of these vectors. Both cases, the classical and the quantum, employ a non-linear activation leading to the output $\sigma$.
\textbf{(c)} Quantum circuit implementation of the quantum perceptron following the work of Tacchino et al. ~\cite{Tacchino2019}. An encoding unitary realizes the input state $\ket{\psi_{\vec{i}}}$ and the processing unitary computes the inner product of the input and weight vectors. The outcome is then written on the ancilla qubit with a multi-controlled NOT gate. Finally, the activation is measured by the readout of the ancilla qubit.
}
\label{Fig1} 
\end{figure}

A quantum analog of the classical perceptron~\cite{Tacchino2019} is depicted in Fig.~\ref{Fig1}b with the corresponding quantum circuit in Fig.~\ref{Fig1}c. In this quantum perceptron the connection between the inputs, outputs and weights is given by the activation function
\begin{equation} \label{qua_act}
     \sigma^{\mu} = \theta \left(\tfrac{1}{m}|\vec{i}^{\mu} \cdot \vec{w}|^2 - \kappa \right) \, ,
\end{equation}
where the non-linearity of the perceptron is realized by the measurement, see Fig.~\ref{Fig1}b. 

In the quantum case the input vector is $\vec{i}^{\mu} = (i^{\mu}_0, \ldots, i^{\mu}_{m-1})^T$ and the weight vector is $\Vec{w} =(w_0, \ldots, w_{m-1})^T$, where $m$ is the dimension of the Hilbert space. The vectors $\vec{i}^{\mu}$ and $\vec{w}$ are encoded in quantum states
\begin{subequations}
\begin{align}
    \ket{\psi_{\vec{i}}} &=  \frac{1}{\sqrt{m}}\sum_{j=0}^{m-1} i_j \ket{j} \, ,\\
    \ket{\phi_{\vec{w}}} &=  \frac{1}{\sqrt{m}}\sum_{j=0}^{m-1} w_j \ket{j} \, ,
\end{align}
\end{subequations}
respectively with the orthonormal basis vectors $ \ket{j} $ form the computational basis and we focused on the case of binary inputs and weights. 
The encoding unitary $U_{\vec{i}}$ prepares the input state, while the processing unitary $V_{\vec{w}}$ computes the inner product between $\vec{i}$ and $\Vec{w}$. The precise definition of the unitaries 
can be found in App.~\ref{app:Tacchino}.
After the encoding and processing step (depicted by two blocks in Fig.~\ref{Fig1}c), a multi-controlled NOT gate is applied between the register and an ancilla qubit. Measuring the ancilla qubit in the computational basis gives $\ket{1}$ with probability $ |\vec{i}^{\mu}\cdot \vec{w}|^2 $, see Fig.~\ref{Fig1}c and App.~\ref{app:Tacchino} for details. \\

\subsection{Gardner's program} The correct choice of the weights results in a desired input-output relation, i.e., a specific mapping between the input $\vec{i}^{\mu}$ and the output $\sigma^\mu$. A learning rule is usually applied to find the correct weights, such as the Hebbian rule~\cite{Hebbrule}. While the Hebbian rule has an appealing simplicity, Gardner, in her works~\cite{Gardner1988a, GD}, was interested in the global properties of the classical perceptron model without specifying the learning rule. She asked the  question: What is the maximum number of input-output patterns that the classical perceptron can realize? Therefore, she considered the relative volume in the space of possible weights, which realizes a given input-output relation. 

\subsection{Storage capacity} 
Following Gardner's work, the storage capacity can be obtained from the fraction of $\vec{w}$-space which correctly reproduces the desired input-output relations normalized to the volume of vectors $\vec{w}$. When increasing the number of patterns, the volume of vectors $\vec{w}$ typically  shrinks, and the relative volume of the weights vanishes. The limit of vanishing relative volume defines the storage capacity of the perceptron \cite{Nishimori2001}.  From the definition of the storage capacity the difference between the classical and quantum perceptrons results from the different dimensionality of the input vectors, which equals $N$ physical inputs for the classical perceptron. In contrast, in the quantum case, the number of inputs equals the dimension of the Hilbert space $m$. Hence, for the classical perceptron we have $\alpha_c = p/N$, while for the quantum perceptron $\alpha_c = p/m$. For the classical perceptron, the storage capacity is known to be $2$ and
was calculated for example in~\cite{Gardner1988a, GD, Muller1995, Nishimori2001}. \\

\subsection{Calculation of the relative volume} 
In the following, we will focus on quantum perceptrons.
The abundance of weights, which lead to desired input-output relations, can be treated by averaging over the weight vectors $\vec{w}$. 
This averaging gives rise to an ensemble of quantum machines, which can be analyzed with statistical physics tools. 
To define a finite volume of weights~\cite{GD} we constrain the weight vector $\vec{w}$. 
Similar to Gardner's work one can consider two types of constraints: spherical weights, i.e.,  $|\vec{w}|^2=m$ and Ising weights $w_i = \pm 1$. The corresponding integration measures~\cite{Muller1995} are 
\begin{subequations}
\begin{align}
\rho_S[\vec{w}] &=\frac{1}{V_{S_0}}\delta\left(|\vec{w}|^2 - m\right) \, , \label{eq:SphericalWeights} \\
\rho_I[\vec{w}]   &=\frac{1}{V_{I_0}} \prod_k [\delta\left(w_{k}-1\right)+\delta\left(w_{k }+1\right) ]\, 
\label{eq:IsingWeights}
\end{align}
\end{subequations}
with the normalization (see Appendix~\ref{sec:Measure})
\begin{subequations} \label{eq:IsingAndSpherical}
\begin{align}
 V_{S_0} &= \int_w \delta ( |\vec{w}|^2 - {m} ) \, , \\
 V_{I_0} &= \int_w \, \prod_k \, \left[\delta\left(w_{k}-1\right)+\delta\left(w_{k}+1\right)\right] \,.
\end{align}
\end{subequations}
Then the relative volume of perceptrons, which fulfill a specific input-output relation, is given by
\begin{align} 
V_M = \int_w \prod_{\mu} \theta \left(  \tfrac{1}{m}|\vec{i}^{\mu}\cdot \vec{w}|^{2} - \kappa  \right) \rho_M[\vec{w}] \label{eq:Volume} \, ,
\end{align}
where the label $M=S$ for the spherical constraint or $M=I$ for the Ising constraint. The threshold $\kappa$ takes values in $[0, m]$ and in the limit  $\kappa \rightarrow 0$ the relative volume
allows us to obtain the maximum storage capacity of the quantum perceptron model~\cite{Nishimori2001, Muller1995}. We 
calculate the relative volume 
using the integral representation of the Heaviside function  
\begin{align}\label{theta}
\theta \left( y- \kappa \right) = \int_{\kappa} ^ {\infty} d\lambda \,
\int_{-\infty}^{\infty} \frac{dx}{2\pi}  e^{i x \left( \lambda - y \right)},
\end{align}
which we insert into Eq.~\eqref{eq:Volume}. In the following 
we outline the calculation of the relative volume for the case of
spherical weights and present the details of the calculation in Sec.~\ref{methods}.
\subsubsection{Spherical weights}  
The distribution of the spherical weights is given in Eq.~\eqref{eq:SphericalWeights} and contains a delta function, which we represent via
\begin{align}
\delta ( |\vec{w}|^2 - m ) = \int_{-\infty}^{\infty} \dfrac{dE}{2\pi} 
e^{ {iE \left( |\vec{w}|^2 - m \right) } } \, .    
\end{align}
Further, we average over the input vector $\vec{i}^{\mu}$ to avoid bias towards specific input vectors. The average with respect to $\vec{i}^{\mu}$ is denoted as $\braket{\braket{\cdot}}$. The expression for the relative volume becomes
\begin{align} 
\braket{\braket{V_S}}
&= \dfrac{ 1 }{V_{S_0}}  \int_{w} \int_{\lambda} \int_{x} \int_{E}  \exp \left[  iE \left( |\vec{w}|^2 - m \right) \right]  \notag \\
&\times \braket{\braket{ \exp \left[ i \sum_{\mu} x^{\mu} \left( \lambda^{\mu}- \tfrac{1}{m}|\vec{i}^{\mu} \cdot \vec{w}|^2 \right) \right]}} \, , \label{eq:averaged_rel_volume}
\end{align}
where the integration measure is given in App.~\ref{sec:Measure}.

Similar to Gardner we make the observation that Eq.~\eqref{eq:averaged_rel_volume} is 
a partition function of a classical spin glass, where
$\braket{\braket{\cdot}}$ is interpreted as a disorder average and $\vec{w}$ is a classical spin variable. 
As for classical spin glasses~\cite{Nishimori2001, Muller1995} we calculate $\braket{\braket{\log V_S}} $ via the replica trick
\begin{align} \label{logVs}
\braket{\braket{\log V_S}}  = \lim_{n \rightarrow 0} \frac{ \braket{\braket{V_S^n}} -1}{n} \, ,
\end{align}
which leads to the replicated variables $\vec{w}^\alpha$, $x^{\alpha}$, $\lambda^{\alpha}$ with the replica index $\alpha \in \{1, \ldots , n\}$.
In addition, we introduce the spin glass order parameter $q^{\alpha\beta}$ and its conjugate $F^{\alpha\beta}$ via the 
integral
\begin{align}\label{order_parameters}
    1 &= \int_{-\infty}^{\infty} dq^{\alpha\beta} \delta \big(  q^{\alpha\beta} -  \tfrac{1}{m} \sum_{k}  w^{\alpha}_k w^{\beta}_k \big)  \notag \\
    &= m \int_{-\infty}^{\infty} dq^{\alpha\beta} \int_{-\infty}^{\infty} \frac{dF^{\alpha\beta}}{2\pi} e^{imF^{\alpha\beta}(q^{\alpha\beta} - \frac{1}{m}\sum_{k}  w^{\alpha}_k w^{\beta}_k)} \, 
\end{align}
with $\alpha<\beta$. This identity is also referred to as Hubbard-Stratonovich transformation, see~\cite{SK,Parisibook} for details.

\subsubsection{Ising inputs} 
In the next step, we perform the average over the inputs and assume
small fluctuations of  $x^{\alpha}$, which leads to
\begin{align} \label{eq:RelativeVolume}
\braket{\braket{V_S^n}} = \dfrac{ 1 }{V^{n}_{S_0}} 
\int_F  \int_q \int_E e^{mG}, 
\end{align}
with integration measure given in App.~\ref{sec:Measure}
and where we introduced the effective potential
\begin{align} \label{eff_potential}
G\!&=\! \alpha G_1 [q^{\alpha \beta}] + G_2 [ E^{\alpha},F^{\alpha \beta}]  - i \sum_{\alpha} E^{\alpha} + i \sum_{\alpha <\beta} F^{\alpha \beta} q^{\alpha \beta} 
\end{align}
with the storage capacity $\alpha$ and the two contributions
\begin{align}
& G_1[q^{\alpha \beta}] = \log 
\int_{-\infty}^{\infty} \prod_{\alpha} \frac{dx^{\alpha}}{2\pi}  \int_{\kappa} ^ {\infty} \prod_{\alpha} d\lambda^{\alpha} \notag \\
&\times \exp \left(i \sum_{\alpha} x^{\alpha} \lambda^{\alpha}-\frac{1}{2} \sum_{\alpha}\left(x^{\alpha}\right)^{2}-\sum_{\alpha < \beta} (q^{\alpha \beta})^{2} x^{\alpha} x^{\beta}\right)  \label{G1-f}
\end{align}
and
\begin{align}
&G_2[E^{\alpha},F^{\alpha \beta}] = \log \int_{-\infty}^{\infty}  \prod_{ \alpha} dw^{\alpha}  \notag \\
&\times \exp \left(i \sum_{\alpha} E^{\alpha} (w^{\alpha})^2-i \sum_{\alpha < \beta} F^{\alpha \beta} w^{\alpha} w^{\beta}\right) \, \label{G2-f} \,.
\end{align}
Comparing the integrals for the effective potential reveals a quadratic dependency on $q^{\alpha\beta}$ for the quantum model and a linear dependence on $q^{\alpha\beta}$ for the classical model within the exponents.
The non-linear dependence in the quantum case is a consequence of the measuring process, which involves the modulus square. 

\begin{figure}
\includegraphics[width=\columnwidth]{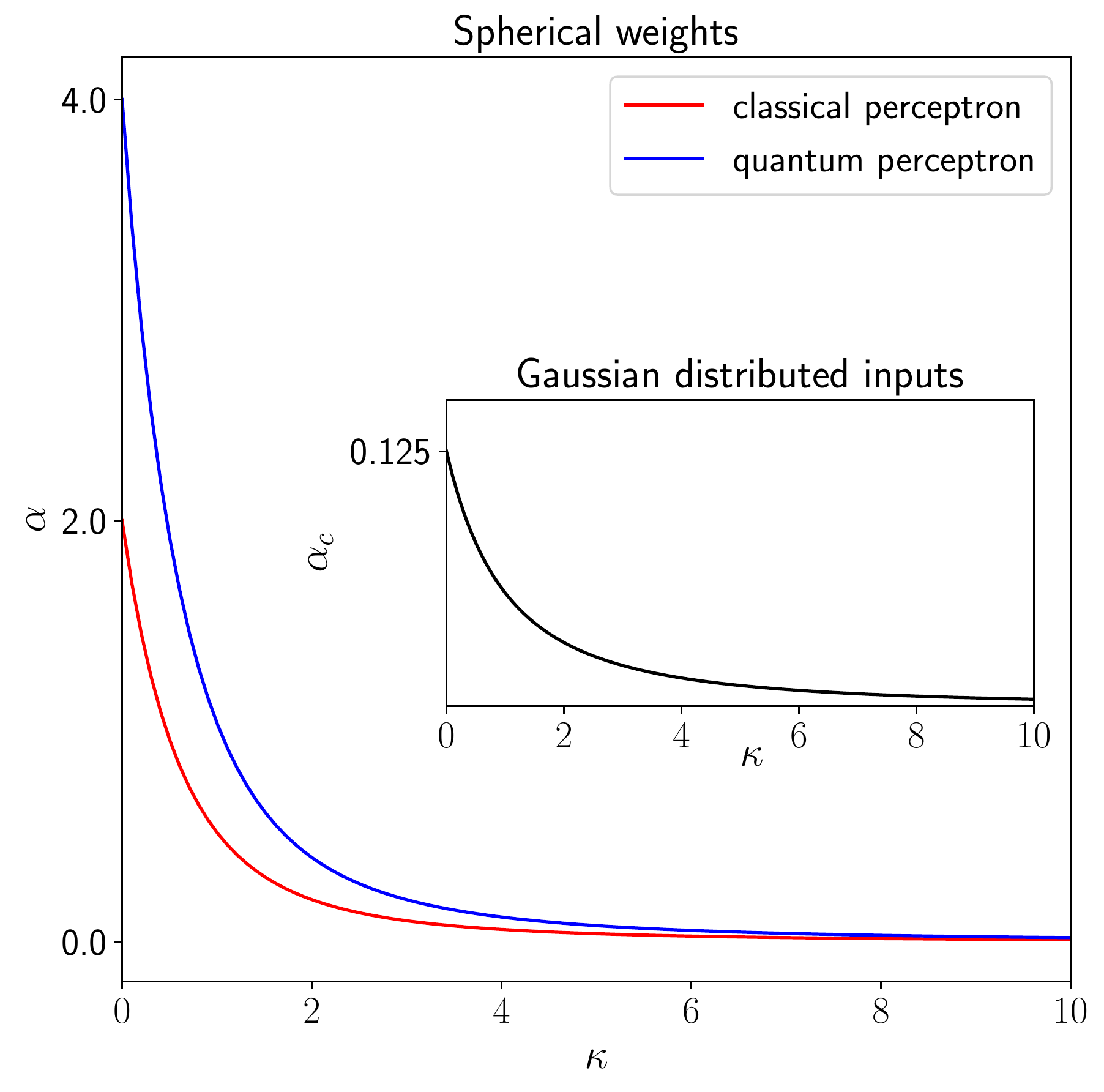}
\caption{\textbf{Storage properties of perceptrons for spherical weights:} Storage capacity for the classical (red) and quantum (blue) perceptrons as a function of the threshold $\kappa$. For $\kappa \approx 0$, the storage capacity has a maximum, whereas the storage capacity decays for $\kappa \gg 1$. In the inset, we plot the critical storage capacity as a function of $\kappa$ for Gaussian distributed inputs.}
\label{alpha-kappa} 
\end{figure}

After the Hubbard-Stratonovich transformation we perform a saddle-point approximation for large $m$ and assume replica symmetry, i.e.,   
\begin{equation}
    q^{\alpha \beta} = q, F^{\alpha \beta} = F, E^{\alpha} = E \, .
\end{equation}
The saddle-point equations are
\begin{equation}
   \frac{ \partial G}{ \partial E} = \frac{ \partial G}{ \partial F} = \frac{ \partial G}{ \partial q} = 0 \, ,
\end{equation}
which we solve and subsequently perform the limit $n \rightarrow 0$.
Taking the derivative of $G$ with respect to $q$ and analyzing the limits $q\rightarrow1$ and $\kappa \rightarrow 0$ leads to the maximum critical storage capacity of 
\begin{align}\label{eq_cap}
    \alpha_{c,\text{max}}= 4.
\end{align}
Additionally, the saddle-point approximation allows us  to study the critical storage
capacity $\alpha_{c}$ as a function of the threshold $\kappa$, which we depict in Fig.~\ref{alpha-kappa}.
\\

\subsubsection{Gaussian distributed inputs} 
In this section, we discuss the case of inputs distributed according to a Gaussian normal distribution. The weights in turn are distributed according to Eq.~\eqref{eq:SphericalWeights}. We apply the replica trick Eq.~\eqref{logVs}, introduce the order parameters as in Eq.~\eqref{order_parameters} and average over the inputs to calculate the effective potential $G$. A comparison to Eq.~\eqref{eq:RelativeVolume} reveals that only the expression for $G_1$
changes
\begin{align}
& G_1 [q^{\alpha \beta}] = \log 
\int_{-\infty}^{\infty} \prod_{\alpha} \frac{dx^{\alpha}}{2\pi}  \int_{\kappa} ^ {\infty} \prod_{\alpha} d\lambda^{\alpha}  \notag \\
&\times\exp \left[ i \sum_{\alpha} x^{\alpha} \lambda^{\alpha} - \log \det \left(1+2i \hat A \right) \right] \label{G1-G} \, ,
\end{align}
where we introduce the matrix $A$ later in Eq.~\eqref{eq:MatrixA}.
Next, we assume replica symmetry, i.e., 
\begin{equation}
    q^{ab} = q, F^{ab} = iF,
\end{equation}
and the saddle-point equations  are
\begin{equation}
\frac{ \partial G}{ \partial F} = \frac{ \partial G}{ \partial q} = 0,
\end{equation}
which we solve and subsequently perform the limit $n \rightarrow 0$.
The saddle-point equation given by the derivative with respect to $q$ leads to
\begin{align}
\alpha \left( 2 + \kappa \right)^2 q = \dfrac{q}{2\left( 1 - q \right)^2 }.
\end{align}
This equation has one trivial solution $q=0$ and one non-trivial in $0<q<1$. The non-trivial solution exists if and only if
\begin{align}
2\alpha \left( 2 + \kappa \right)^2 \ge 1. 
\end{align}
For $\alpha\le \alpha_c=(1/2)(2+\kappa)^2$ the solution is trivial, and the logarithm of the relative volume is close to one, it is proportional to $\alpha\kappa/2$, and for $\kappa=0$ all perceptrons are activated. Above $\alpha_c$, the solution is non-zero, and the volume shrinks faster than exponentially with $m$. 
We plot $\alpha_c$  in Fig.~\ref{alpha-kappa} for different values of $\kappa$ and observe $\alpha_c\to 1/8$ for $\kappa=0$. The phase transition has different character in comparison to Gardner's work~\cite{Gardner1988a}. In her work, the volume decreases exponentially with $m$ below the critical $\alpha_c$ (where $q<1$), and strictly shrinks to zero above the critical $\alpha_c$ (where $q=1$). In our work, the volume is close to one below $\alpha_c$, and it starts to shrink exponentially with $m$ above.\\

\subsubsection{Ising weights} 
\begin{figure}
\includegraphics[scale=0.3]{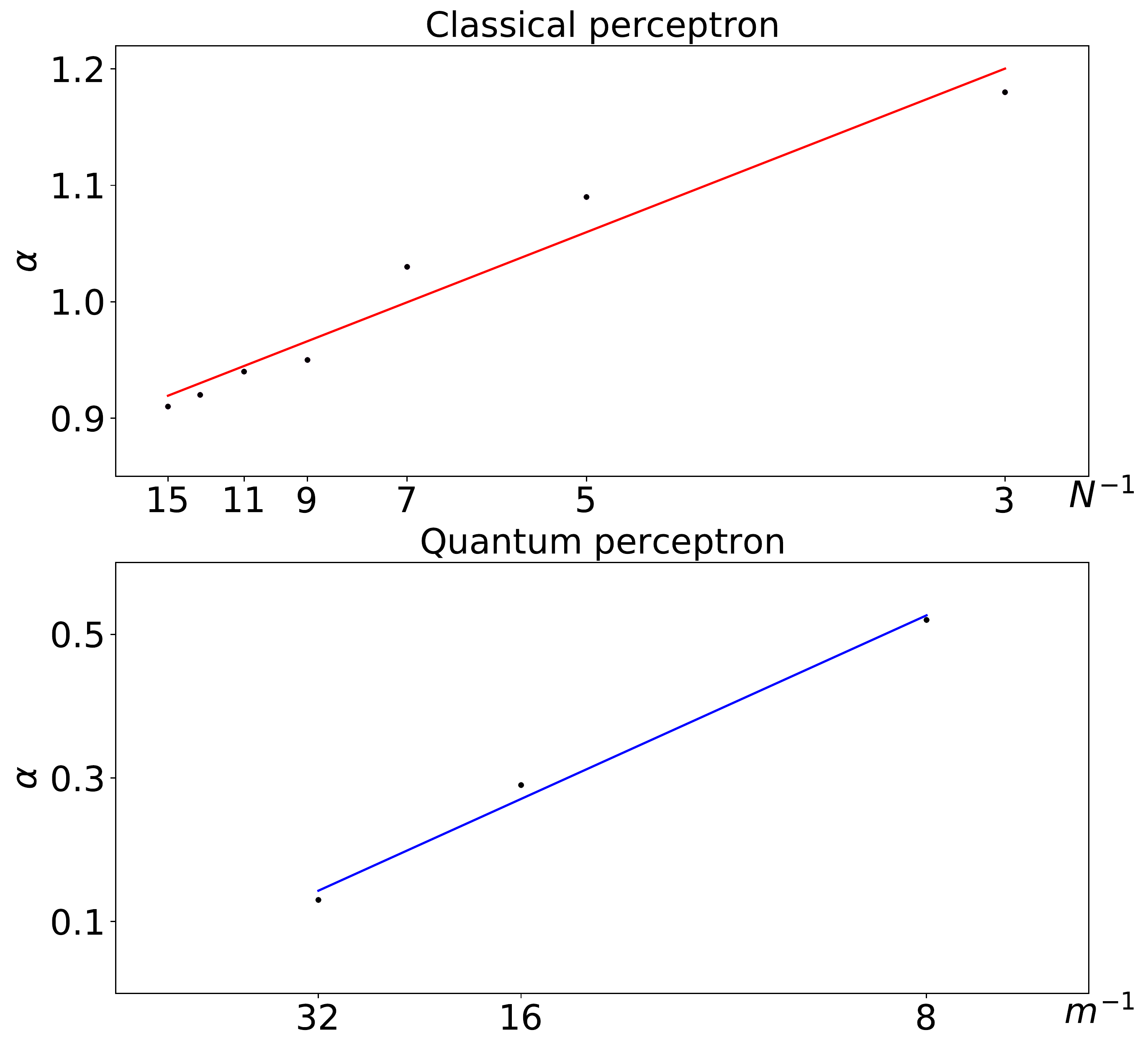} 
\caption{\textbf{Storage properties of perceptrons for the Ising weights} Storage capacity for the classical (red) and quantum (blue) Ising perceptrons as a function of the number of inputs $N^{-1}$ and $m^{-1}$, respectively. The dots are the result of the Monte Carlo simulations. The intersection of the lines with the $y$-axis gives the storage capacity $ 0.86 \pm 0.01$ for the classical and $ 0.010 \pm 0.005$ for the quantum perceptron in the limits $N, m~\rightarrow~\infty$. }
\label{MC}
\end{figure}

In the classical case the Ising weights were treated for example in~\cite{Nishimori2001, Muller1995, GD}. Here, 
we use the Ising weights for the
quantum case and employ Eq.~\eqref{eq:IsingWeights} and Eq.~\eqref{eq:IsingAndSpherical} for the integration measure and normalization of the volume, respectively. 
We apply the replica trick Eq.~\eqref{logVs}, introduce the order parameters Eq.~\eqref{order_parameters} and average over the inputs to calculate the effective potential. The contribution $G_1$ is the same as Eq.~\eqref{G2-f}, while $G_2$ becomes
\begin{equation}\label{G2-Ising}
G_2[F^{\alpha \beta}] = \log \sum_{\{ w^{\alpha} =  \pm 1 \}} \exp \left( \sum_{\alpha < \beta} F^{\alpha \beta} w^{\alpha} w^{\beta} \right).
\end{equation}
We assume replica symmetry, i.e., 
\begin{equation}
    q^{ab} = q, F^{ab} = iF,
\end{equation}
and the saddle-point equations are
\begin{equation}
\frac{ \partial G}{ \partial F} = \frac{ \partial G}{ \partial q} = 0.
\end{equation}
Solving the saddle point equations 
in the limit  for $q\rightarrow1$ we conclude that  the storage capacity is
\begin{align}\label{eq_cap}
    \alpha_c (\kappa) =\frac{4}{\pi}\left[\int_{-\kappa}^{\infty} Dy (\kappa + y)^2 \right]^{-1}, 
\end{align}
where we used the abbreviation
\begin{align}
    \int_{-\infty}^{\infty} D y\, =\sqrt{\frac{1}{2 \pi}} \int_{-\infty}^{\infty} d y\,\, e^{-\frac{y^{2}}{2}} \,.
    \label{eq:abb}
\end{align}
In addition, we present the results of a Monte Carlo simulation in Fig.~\ref{MC}, which shows that as $m \rightarrow \infty$, the storage capacity approximately behaves as $\alpha\rightarrow 0$  (see App.~\ref{app:MC} for details). We interpret this discrepancy as the necessity for replica symmetry breaking. This analysis, however, goes beyond the scope of this paper. 

\section{Discussion \label{sec:Discussion}}
In this work, we calculate the storage capacity of a quantum perceptron, which was proposed in a recent work~\cite{Tacchino2019}. Following the seminal works of Gardner~\cite{Gardner1988a, GD}, we use statistical physics techniques to calculate the storage capacity of this perceptron. In particular, we interpret this quantum perceptron as a classical perceptron on an extended input space with a different activation function, see Fig.~\ref{Fig1}. This interpretation allows us to calculate the storage capacity of a quantum perceptron by computing the relative volume of quantum perceptrons which fulfill a given input-output relation.

To handle the multitude of inputs and learning rules, we integrate over the 
input and the weights. Formally, this averaging
over input and weights maps the calculation of the relative volume to the partition
function of a classical spin glass problem. Similar to problems
in classical spin glass theory, we compute the logarithm of the partition function using the replica trick~\cite{Nishimori2001, Muller1995}. Further, by using the large $m$ expansion, we can determine the storage capacity $\alpha_c$ in leading order, which is the ratio of the stored patterns $p$ over the computational resources $m$. Notably, the techniques presented here are applicable to other quantum architectures.

 Given the model of Fig.~\ref{Fig1}c, we obtain a maximal critical storage capacity of { $\alpha_{c,\text{max}} = 4$} for the spherical weights, see Fig.~\ref{alpha-kappa}. To put these results in perspective, we compare them with the classical perceptron. In the classical case, the maximal storage capacity is {$\alpha_{c,\text{max}} = 2$}, see Fig.~\ref{alpha-kappa}.  With our definition of the storage capacity, the classical and quantum perceptron lead to the same order of magnitude when $\kappa$ goes to zero. However, we  emphasize that the input vector of the perceptron is determined by the number of inputs $N$ in the classical case, but equals the dimension of the Hilbert space $m$ in the quantum case. In particular, the number of classical inputs and the dimension of Hilbert space are connected via $m=2^N$. Our results on the storage capacity are also in accordance with recent works~\cite{ANDRECUT2003, VENTURA, wright2019capacity, lewenstein2020storage}. 
 
 For Gaussian distributed inputs, the performance of the quantum perceptron is quite different from the classical perceptron in accordance with a related work~\cite{new_related_work}. In the classical case, the relative volume shrinks exponentially with $m$  below the critical capacity, and shrinks suddenly to zero above $\alpha_c$, see \cite{MappingCG}. In the present study, the volume shrinks exponentially with $m$, but the rate of shrinking changes from below ("easy learning" phase) to above $\alpha_{c}$ ("hard learning" phase). The maximum storage capacity of the "easy learning" phase is $0.125$.
 
For a classical perceptron with Ising weights, analytical calculations suggest that $\alpha_{\text{max}} = 0.83$ \cite{krauth}, while our Monte Carlo simulations gives $\alpha_{\text{max}} = 0.86 \pm 0.01$~\cite{Gardner:unfinished}. In contrast to the classical case, the relative volume of interactions goes approximately to zero for quantum perceptrons with Ising weights according to our MC simulations. This is in accordance with results from the related work in~\cite{Tacchino2019}, where they explore the learning capability of quantum perceptrons with Ising weights. They show that for each pattern stored a different set of weights is required. This result already suggests that it is hard to choose a weight that can store many patterns.

For future work, it would be highly important for practical applications to explore the storage properties of the quantum perceptron with correlated inputs or input-output patterns.
In a previous analysis of quantum perceptrons~\cite{QP_Lewenstein}, one distinguished between three different phases an \textit{ignorant \; phase}, a \textit{random \; phase}, and \textit{learning \; phase}, and it would be interesting to detect these phases in the quantum perceptron architecture of~\cite{Tacchino2019}.
Future studies should also investigate the storage capacity away from $q\approx 1$ and the dependency on $\kappa$. Also, it will be essential to include corrections to the large $m$ expansion and study the stability of the replica symmetric saddle point solution \cite{Parisibook}.
Finally, an exciting continuation of this work would be to consider other architectures of quantum perceptrons \cite{Schuld-perceptron, Erik_perceptron, gratsea2021exploring}, e.g., qudit based platforms~\cite{Kasper2020, Weggemans2021, Ringbauer2021} and analyze them with the tools presented in this work. 

\section{Methods\label{methods}}
In this section, we elaborate on the computational details presented in Sec.~\ref{sec:Results}, i.e., the averaging over the inputs, the calculation
of the effective potential, and the saddle-point approximation.\\

\subsection{Averaging over the input patterns}
We perform the average $\braket{\braket{\cdot}}$ and assume weak correlations between the weights. Then we can approximate
\begin{align}
    & \braket{\braket{\prod_{\alpha,\mu} e^{-\frac{i}{m} x^{\alpha}_{\mu} |\vec{i}^{\mu}\cdot \vec{w}^{\alpha}|^{2}}}}
     \approx \prod_{\mu,k,l}  \cos \Big( \tfrac{1}{m} \sum_{\alpha} x^{\alpha}_{\mu} w_k^{\alpha} w_l^{\alpha}    \Big) \! \, ,
     \label{eq:approx}
\end{align}
see~\cite{Kohring1990} for details.
Using Eq.~\eqref{eq:approx} the relative volume becomes
\begin{align} \label{eq:AveragedVolume}
    & \braket{\braket{V_S}}
= \dfrac{ 1 }{V^n_{S_0}}  \int_{w} \int_{\lambda} \int_{x} \int_{E}  \int_{q} \int_{F}  \notag \\
    & \times \exp \Big[ i \sum_{{\alpha}, \mu} x^{\alpha}_{\mu} \lambda^{\alpha}_{\mu}  + \sum_{\mu,k,l} \log \cos \Big( \tfrac{1}{m} \sum_{\alpha} x^{\alpha}_{\mu} w_k^{\alpha} w_l^{\alpha}    \Big) \Big] \, \notag \\
    & \times \exp \left[  iE^{\alpha} \left( |\vec{w}^{\alpha}|^2 - m \right) +imF^{\alpha\beta}(q^{\alpha\beta} - \frac{1}{m}\sum_{k}  w^{\alpha}_k w^{\beta}_k) \right] \,.
\end{align}
Next, we use the approximation $\log \cos{x} \approx -x^2/2$ and use Eq.~\eqref{order_parameters}.
In addition, we employ that the integral
\begin{align}
&I = \int_{-\infty}^{\infty} \prod_{\alpha,\mu} \frac{dx_{\mu}^{\alpha}}{2\pi} 
\int_{\kappa} ^ {\infty} \prod_{\alpha,\mu} d\lambda_{\mu}^{\alpha} \notag \\
&\times \exp \left(i \sum_{\alpha,\mu} x_{\mu}^{\alpha} \lambda_{\mu}^{\alpha}-\frac{1}{2} \sum_{\mu,\alpha}\left(x_{\mu}^{\alpha}\right)^{2}-\sum_{\alpha < \beta, \mu} (q^{\alpha \beta})^{2} x_{\mu}^{\alpha} x_{\mu}^{\beta}\right)  
\end{align}
factorizes according to
\begin{align}
&I = \Bigg[ \int_{-\infty}^{\infty} \prod_{\alpha} \frac{dx^{\alpha}}{2\pi} 
\int_{\kappa} ^ {\infty} \prod_{\alpha} d\lambda^{\alpha}  \notag \\
&  \times e^{  i \sum_{\alpha} x^{\alpha} \lambda^{\alpha}-\frac{1}{2} \sum_{\alpha}\left(x^{\alpha}\right)^{2}-\sum_{\alpha < \beta} (q^{\alpha \beta})^{2} x^{\alpha} x^{\beta} }\Bigg]^p  \, ,
\end{align}
which leads to Eq.~\eqref{eq:RelativeVolume} - \eqref{G2-f}.

\subsection{Calculation of $G_1$} We assume replica symmetry of $q^{ \alpha \beta }$ and after the integration over $x^{\mu}$ we have 
\begin{align}
    \lim_{n\rightarrow 0} \frac{1}{n}G_1[q] = \int_{-\infty}^{\infty} Dy \log L(y) \, ,
\end{align}
where we used the abbreviation Eq.~\eqref{eq:abb} and introduce
\begin{align}\label{eq:DefL_y}
    L(y) = 2 \sqrt{\pi} \operatorname{Erfc}\left[\frac{\kappa+y q}{\sqrt{2(1-q^2)}}\right]\,.
\end{align}

The function $L(y)$ is the main object that distinguishes the 
classical and the quantum perceptron. In the classical case we have
\begin{align}\label{eq:DefL_y}
    L(y) = 2 \sqrt{\pi} \operatorname{Erfc}\left[\frac{\kappa+y q}{\sqrt{(1-q)}}\right].
\end{align}
In the quantum case $L(y)$ depends on $q^2$, since we are dealing with squared scalar products, which leads to an additional factor of two in the denominator of $L(y)$; this factor will then be responsible for the increase of the storage capacity for the quantum case in comparison to the classical case.

\subsection{Calculation of $G_2$ for spherical weights} We also assume replica symmetry of $E^{\alpha}$ and $F^{ \alpha \beta }$ and perform the multi-dimensional Gaussian integral in Eq.~\eqref{G2-f} resulting in
\begin{align}
    {G_{2}[E, F]}= \log \left[(2 \pi i)^{n / 2}(\operatorname{det} M)^{-1 / 2}\right],
\end{align}
where we introduced the matrix
\begin{align}
    M^{ab} =  (2E+F) \delta^{ab} - F  \,.
\end{align}
The  matrix $M$ has $n-1$ degenerate eigenvalues $\Lambda_1 = \ldots = \Lambda_{n-1} = 2E + F$ and one non-degenerate eigenvalue $\Lambda_n = 2E - (n-1)F$ such that the determinant of the matrix $M$ becomes
\begin{align}
\log \operatorname{det} M =(n-1)  \log (2E + F)+\log [2E - (n-1)F]\,.
\end{align}

\subsection{Saddle-point equations of $G$ for spherical weights}
Since $G_1$ does not depend on $E$ and $F$ the saddle-point equations with respect to $E$ and $F$ are 
\begin{subequations}
\begin{align}
0&=\frac{1}{n}\frac{\partial G}{\partial E}=-i+\frac{1}{n}\frac{\partial G_{2}}{\partial E}, \\
0&=\frac{1}{n}\frac{\partial G}{\partial F}= \frac{i}{2} (n-1) q+\frac{1}{n}\frac{\partial G_{2}}{\partial F} \, ,
\end{align}
\end{subequations}
with
\begin{subequations}
\begin{align}
    \frac{1}{n}\frac{\partial G_{2}}{\partial E} &= \frac{(n-1) F+2 E(n-2)}{2(2 E+F)(-F n+2 E+F)},  \\
    \frac{1}{n}\frac{\partial G_{2}}{\partial F} &= \frac{(n-1) F}{2(2 E+F)(-F n+2 E+F)}.
\end{align}
\end{subequations}
Performing the limit $n \rightarrow 0$ and solving for $E$ and $F$ results in
\begin{subequations}
\begin{align}
E&= \frac{i(1-2 q)}{2(1-q)^{2}}, \\
F&= \frac{iq}{(1-q)^{2}}.
\end{align}
\end{subequations}
Further, we define the effective potential
\begin{align}
    g = \lim_{n \rightarrow 0} \frac{1}{n} G 
\end{align}
and insert the solution of the saddle point equation into $G$. 
As a result, we obtain 
\begin{align}\label{G_fig4}
    g &=  \alpha\int_{-\infty}^{\infty} Dy  \log L(y)  + \frac{1}{2} \log \left( 1-q \right) +     \frac{1}{2 \left( 1-q \right) } \, 
\end{align}
plus constant terms independent of $q$. We can interpret the logarithm of $G$ as a free energy, which is a regular function of $0\le q < 1$, but has a singularity at $q=1$. Employing the  asymptotic expansion of $\operatorname{Erfc}(x) \approx \sqrt{\pi} x^{-1} e^{-x^2}\theta(x)$ for $x\rightarrow \infty$, we see that
\begin{align}\label{G_final}
    g &\simeq -\frac{\alpha}{2(1-q)}\int_{-\infty}^{\infty} Dy (\kappa +y)^2+ \frac{1}{2 \left( 1-q \right) } \, ,
\end{align}
has two singular terms as $q\to 1$. 

If the second term dominates, it repulses the saddle point solution for $q$ from being close to one. In this scenario, the order parameter $q$ is finite as well as the function $G$. The relative volume shrinks in this regime exponentially with $m$. If the first term dominates above $\alpha_c$, the minimum of the free energy immediately shifts to $q=1$, where $g \rightarrow - \infty$, so that the relative volume shrinks immediately to zero.

\begin{figure}
\includegraphics[width=\columnwidth]{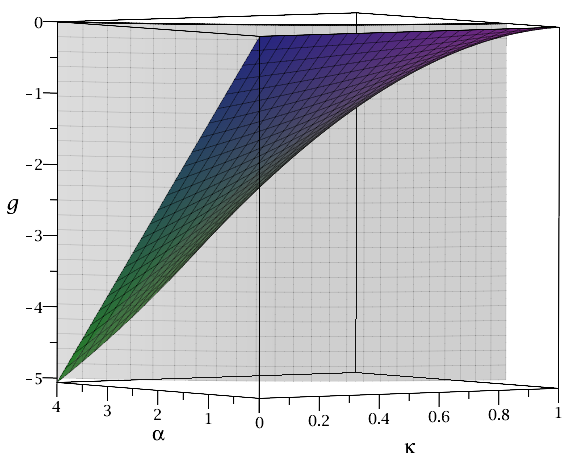}
\caption{\textbf{The effective potential for the spherical weights} We plot the effective potential $g$ of Eq.~\eqref{G_fig4}. It changes from finite negative values below the $\alpha_c$ curve to $-\infty$ (represented by the grey area) above. The black line shows the $\alpha_c$ curve given by Eq.~\eqref{alpha_critical}.}
\label{fig4} 
\end{figure}

We plot the effective potential as a function of $\alpha$ and $\kappa$ in Fig.~\ref{fig4}.
This analysis implies that the critical value of the storage capacity is
\begin{align}\label{alpha_critical}
    \frac{\alpha_{c}}{2} \int_{-\kappa}^{\infty} Dy  \left( \kappa + y \right)^{2}   =     1.
\end{align}
Performing the integral leads to the maximal critical storage capacity of $\alpha_{c,\text{max}} =  4$ for $\kappa \rightarrow 0$.\\

\subsection{Calculation of $G_1$ for Gaussian distributed inputs}
First, we define the matrix

\begin{align}
A_{kk'} = \dfrac{1}{n} \sum_{\alpha} x^{\alpha} w_k^{\alpha} w_{k'}^{\alpha} \, ,
\label{eq:MatrixA}
\end{align}
which is spanned by the vectors $w^{\alpha}$ 
We can write its nontrivial eigenvectors as combinations of $w^{\alpha}$. The eigenvalue equation is
\begin{align}
    \frac{1}{n} \sum_{\alpha,k'} x^{\alpha} w_{k}^{\alpha} w_{k^{\prime}}^{\alpha} \sum_\beta c_{\beta} w_{k^{\prime}}^{\beta} = \Lambda \sum_{\alpha} c_{\alpha} w_{k}^{\alpha} \,.
\end{align}
All other eigenvectors of $1 + 2i\hat{A}$ (orthogonal to the vectors $w^{\alpha}$) are trivial -- they correspond to eigenvalues 1 and do not contribute to the $\log\det$.
Comparing coefficients, using the definition of $q^{\alpha \beta}$ and assuming replica symmetry $q^{\alpha\beta} = q$ for $\alpha \ne \beta$, leads to a closed equation
\begin{align}\label{eigenvalues}
1 = \sum_{a} \dfrac{x^{\alpha} q }{  \Lambda - x^{\alpha} \left( 1- q \right)}
\end{align}
for eigenvalues $\Lambda$ of $\hat A$.
Using the eigenvalues of $\hat{A}$ we can rewrite $G_1$ as
\begin{align}
G_1 [q] = \log \int \prod_\alpha d\lambda^{\alpha} \int \prod_\alpha dx^{\alpha} \dfrac{ \exp \left( i \sum_{\alpha} x^{\alpha} \lambda^{\alpha}\right) }{ \prod_n\left(1+2i \Lambda_n\right)}.
\end{align}
In order to rewrite the product of eigenvalues 
we transform the self-consistent equation for the eigenvalues $\Lambda$
into the characteristic polynomial of $\hat{A}$
to define the function
\begin{align}
    W(\Lambda,x)&=\prod_{\alpha}\left[\Lambda-(1-q) x^{\alpha}\right] \notag \\
    &-\sum_{\alpha} q x^{\alpha} \prod_{\alpha \neq \beta}\left[\Lambda-(1-q) x^{\beta}\right] \, .
\end{align}
Next, we introduce the auxiliary quantity 
\begin{align}
L(\epsilon)=\log\det(1+2i\epsilon\hat A)= \sum_n\log(1+2i\epsilon\Lambda_n) \, ,
    \end{align}
where we are interested in the value of $L(1)$. Differentiating with respect to $\epsilon$ we obtain
\begin{align}
 \dfrac{dL}{d\epsilon} = \dfrac{n}{\epsilon} - \dfrac{1}{\epsilon} \sum_{\Lambda} \dfrac{1}{1+2i\Lambda \epsilon} \,.
\end{align}
The sum can be rewritten by using Cauchy's theorem and employing an appropriate contour $\mathcal{C}$. Using this integral representation for 
the sum we obtain
\begin{align}
 \dfrac{dL}{d\epsilon} =  \dfrac{n}{\epsilon} + \frac{d }{d\epsilon}\log W\left( i/(2\epsilon), x \right) \, .
\end{align}
Integrating  $\epsilon $ from 0 to $1$ we get
\begin{align}
L(1) = \log W\left( i/2, x\right)
\end{align}
since $L(\epsilon)$ goes to zero for $\epsilon \rightarrow 0$.
Expanding $\Lambda$ in $q$, i.e., treating $q$ as a perturbation, $G_1$ becomes
\begin{align}
& G_1 [q] = \log \int \prod_\alpha d\lambda^{\alpha} \int \prod_\alpha dx^{\alpha} \exp \left( i \sum_{\alpha} x^{\alpha} \lambda^{\alpha} \right) \notag \\
&\times \dfrac{1}{ \prod_{\alpha} \left( \dfrac{i}{2} - x_{\alpha} \right)} \left[ 1 + \sum_{\alpha, \beta} \dfrac{q^2 x^{\alpha} x^{\beta}}{\left( \dfrac{i}{2} - x_{\alpha} \right) \left( \dfrac{i}{2} - x_{\beta} \right)} \right]. \label{G1-Gf}
\end{align}
Performing the integration over $x^{\alpha}$ and $\lambda^{\alpha}$ 
gives
\begin{align}
G_1 (q) = \dfrac{\kappa n}{2} - \dfrac{1}{2} n q^2 \left( 2+ \kappa \right)^2.
\end{align}
Due to the perturbative expansion in $q$ the function $G_1(q)$ does not exhibit any singularity at $q=1$ which will affect the nature of the phase transition.\\

\subsection{Saddle-point equations of $G$ for Gaussian distributed inputs}
The effective potential of Eq.~\eqref{eff_potential} becomes
\begin{multline}\label{G_general}
g = \alpha \left[ \dfrac{ \kappa}{2} - \dfrac{1}{2}  q^2 ( 2+ \kappa)^2  \right] + \dfrac{1}{2} \log \left( 1-q\right) + \dfrac{1}{2 \left( 1-q\right)},
\end{multline}
plus constant terms independent of $q$. This function does contain a singular term $(1-q)^{-1}$, which repulses the saddle point solutions for the minimal value away from $q=1$. Indeed, taking the derivative of $G$ with respect to $q$ gives
\begin{align}
\alpha \left( 2 + \kappa \right)^2 q = \dfrac{q}{2\left( 1 - q \right)^2 }.
\end{align}
This equation has a trivial solution $q=0$ for which 
\begin{align}\label{G_small}
g =  (\alpha \kappa)/2,
\end{align}
and a non-trivial solution, which exists for $2\alpha \left( 2 + \kappa \right)^2 \ge 1$. The critical value of the storage capacity is given by (see App.~\ref{appendix_D})
\begin{align}\label{alpha_critical_gaussian}
    \alpha_c= \dfrac{1}{2(2+\kappa)^2}.
\end{align}

\begin{figure}
\includegraphics[width=\columnwidth]{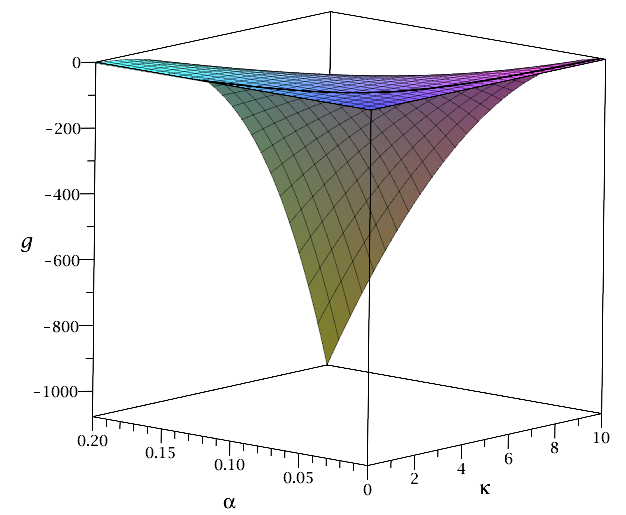}
\caption{\textbf{The effective potential of Gaussian distributed inputs} We plot the effective potential $g$. It changes from small values close to zero of Eq.~\eqref{G_small} below the $\alpha_c$ curve to larger negative values of Eq.~\eqref{G_general} above. The black line shows the $\alpha_c$ curve given by Eq.~\eqref{alpha_critical_gaussian}.}
\label{fig5} 
\end{figure}

Note that the phase transition in this case has a different character: for both solutions $g$ takes a finite value, but it changes from $(\alpha \kappa) / 2$ in the "easy to learn" phase to a larger values in the  "hard to learn phase", see Fig.~\ref{fig5}. This behavior might be the result of expansion in $q$ that we used to obtain the effective potential. Further, the 
effective potential might be the fist term of expansion of $(1-q^2)^{-1}$ and we 
discuss this idea in App.~\ref{appendix_D}, which will bring us back to the volume shrinking phase transition à la Gardner.

\subsection{\it{Calculation of $G_2$ for Ising weights}} Assuming replica symmetry of $F^{ab}$ and averaging over the binary weights, Eq.~\eqref{G2-Ising} becomes
\begin{equation}
G_2 =  - \dfrac{1}{2} Fn + n \int_{-\infty}^{\infty} Dz \log \left[ 2 \cosh ( z \sqrt{F} ) \right],
\end{equation}
as in the classical case~\cite{krauth}.

\subsection{\it{Saddle-point equations of $G$ for Ising weights}}
The effective potential with replica symmetry of $q^{ab}$ and $F^{ab}$ becomes
\begin{align}\label{G_final}
    g = 
    \alpha \int_{-\infty}^{\infty} Dy  \log L(y) + R(F, q) 
\end{align}
with
\begin{equation}
    R(F, q) = -\frac{1}{2} F \left( 1 - q \right) + \int_{-\infty}^{\infty} Dz \log \left[ 2 \cosh \left( z \sqrt{F} \right) \right]. 
\end{equation}
Then, the saddle-point equation with respect to $F$ is
\begin{align}\label{F}
      - \dfrac{1}{2} \left( 1 - q \right)  +  \int_{-\infty}^{\infty} Dz \dfrac{z}{2\sqrt{F}}  \tanh \left( z \sqrt{F} \right)  = 0,
\end{align}
which is similar to equation obtained  for the classical perceptron with binary weights~\cite{GD}, where it  was argued that the solution with   $F \rightarrow \infty$ as $q \rightarrow 1$ is invalid. Instead, for the correct solution of the classical perceptron problem, replica-symmetry breaking must be taken into account. 

Here, we analyze results for replica symmetric case and compare them with Monte Carlo simulations.  For the quantum perceptron, $F$ is a well defined function of $q$,  and it tends to infinity  as $q \rightarrow 1$. The last equation can be solved  
\begin{align}\label{F}
     \sqrt{F} \approx  \sqrt{\dfrac{2}{\pi}}\dfrac{1}{ \left( 1 - q \right)} ,
\end{align}
where ${\rm sign}(x)=x/|x|$ and we approximated the $\rm tanh$ by
$\rm sign(x)$.

Comparing the leading terms when $q \rightarrow 1$, we arrive at
\begin{align}\label{G_final1}
    g &\approx \dfrac{1}{ \left( 1 - q \right)}\left[
    -\frac{\alpha}{4} \int_{-\kappa}^{\infty} Dy (\kappa + y)^2  +\frac{1}{\pi} \right].
\end{align}
In this way we obtain the critical value
\begin{align}
    \alpha_c (\kappa) =\frac{4}{\pi} \left[ \int_{-\kappa}^{\infty} Dy (\kappa + y)^2 \right]^{-1},
\end{align}
which implies the maximal value of $\alpha_c (0)=8/\pi$. In contrast the MC simulations suggest that $\alpha_c(0)\simeq 0$ as illustrated in Fig.~\ref{MC}. As in the classical case, we interpret this discrepancy as the necessity of replica symmetry breaking. 

\section{Acknowledgments}
ICFO group acknowledges support from: ERC AdG NOQIA; Agencia Estatal de Investigación (R\&D project CEX2019-000910-S, funded by MCI/ AEI/10.
13039/501100011033, Plan National FIDEUA PID2019-106901GB-I00, FPI, QUANTERA MAQS PCI2019-111828-2, QUANTERA DYNAMITE PCI2022-132919,  Proyectos de I+D+I “Retos Colaboración” QUSPIN RTC2019-007196-7), MCIN via European Union NextGenerationEU (PRTR);  Fundació Cellex; Fundació Mir-Puig; Generalitat de Catalunya through the European Social Fund FEDER and CERCA program (AGAUR Grant No. 2017 SGR 134, QuantumCAT \ U16-011424, co-funded by ERDF Operational Program of Catalonia 2014-2020); EU Horizon 2020 FET-OPEN OPTOlogic (Grant No 899794); National Science Centre, Poland (Symfonia Grant No. 2016/20/W/ST4/00314); European Union’s Horizon 2020 research and innovation programme under the Marie-Skłodowska-Curie grant agreement No 101029393 (STREDCH) and No 847648  (“La Caixa” Junior Leaders fellowships ID100010434: LCF/BQ/PI19/11690013, LCF/BQ/PI20/11760031,  LCF/BQ/PR20/11770012, LCF/BQ/PR21/11840013); European Union’s Horizon 2020 research and innovation programme under the Marie Skłodowska-Curie grant agreement No 847517.

\section{Data availability}

The data that support the findings of this study are available from the corresponding author upon request.

\section{Author Contributions}

All authors contributed to the design and implementation of the research, to the analysis of the results and to the writing of the manuscript. The Authors declare no Competing Financial or Non-Financial Interests. Correspondence and requests for materials should be addressed to A.G. (gratsea.katerina@gmail.com).

\appendix
\section{Details on the quantum perceptron proposed in~\cite{Tacchino2019} \label{app:Tacchino}}
The first unitary $U_{\vec{i}}$ should fulfill
\begin{align}
\ket{\psi_{\vec{i}}} = U_{\vec{i}} \ket{0}^{\otimes N} \, ,    
\end{align}
and in this way encodes the information on $N$ qubits.
Particularly, any $m \times m$ unitary matrix with the first column being identical with $\vec{i}$ and normalized is a valid candidate for such a unitary.
The information is processed by applying the
second unitary $V_{\vec{w}}$ which fulfills
\begin{align}
V_{\vec{w}}\left|\psi_{\vec{w}}\right\rangle=|1\rangle^{\otimes N}=|m-1\rangle \,.
\end{align}
Applying the unitary $V_{\vec{w}}$ on the encoded state leads to
\begin{align}
\ket{\phi_{\vec{i},\vec{w}}} \equiv V_{\vec{w}} \ket{\psi_{\vec{i}}} = \sum_{j=0}^{m-1} c_{j} \ket{j} \, .
\end{align}
Performing multi-controlled NOT gates with a readout qubit leads to the state
\begin{align}
\ket{\phi_{i,w}} \ket{0} \ag{=} \sum_{j=0}^{m-2} c_{j} \ket{j} \ket{0} + c_{m-1} \ket{m-1} \ket{1} .
\end{align}
As a result, when measuring 1 on the readout qubit, the probability amplitude is
\begin{align} | c_{m-1} |^{2} =  |\vec{i}^{\mu}\cdot \vec{w}|^2.
\end{align}

\section{Abbreviations \label{sec:Measure}}
In this appendix, we summarize the abbreviation used in the main text.
In Eq.~\eqref{eq:IsingAndSpherical} and Eq.~\eqref{eq:Volume} we used 
\begin{align}
    \int_w = \int_{-\infty}^{\infty} \prod_k dw_k \label{eq:abb1} \, ,
\end{align}
and in Eq.~\eqref{eq:averaged_rel_volume} the measure is
\begin{align}
    \int_{w} \int_{\lambda} \int_{x} \int_{E} &= \int_{-\infty}^{\infty} \prod_{k} dw_k \int_{\kappa} ^ {\infty} \prod_{\mu} d\lambda^{\mu} \, \notag \\
    &\times \int_{-\infty}^{\infty} \prod_{\mu} \frac{dx^{\mu}}{2\pi} \int_{-\infty}^{\infty} \dfrac{dE}{2\pi} 
 \label{eq:abb2} \, ,
\end{align}
in Eq.~\eqref{eq:RelativeVolume} the abbreviation means
\begin{align}
    \int_F  \int_q \int_E = \int_{-\infty}^{\infty} \prod_{\alpha < \beta} dq^{\alpha\beta} \int_{-\infty}^{\infty} \prod_{\alpha < \beta} \frac{dF^{\alpha\beta}}{2\pi} \int_{-\infty}^{\infty} \prod_{\alpha} \dfrac{dE^{\alpha}}{2\pi}  \label{eq:abb3} \, ,
\end{align}
and in Eq.~\eqref{eq:AveragedVolume} we used 
\begin{align}
    &\int_{w} \int_{\lambda} \int_{x} \int_{E}  \int_{q} \int_{F} = \int_{-\infty}^{\infty}  \prod_{k, \alpha} dw_k^{\alpha} \int_{\kappa} ^ {\infty} \prod_{\mu,\alpha} d\lambda^{\alpha}_{\mu} \int_{-\infty}^{\infty} \prod_{\alpha,\mu} \frac{dx^{\alpha}_{\mu}}{2\pi} \notag \\ 
    & \times  \int_{-\infty}^{\infty} \prod_{\alpha} \dfrac{dE^{\alpha}}{2\pi} \int_{-\infty}^{\infty} \prod_{\alpha < \beta} dq^{\alpha\beta} \int_{-\infty}^{\infty} \prod_{\alpha < \beta} \frac{dF^{\alpha\beta}}{2\pi} \label{eq:abb4} \,.
\end{align}

\section{Monte Carlo simulation \label{app:MC}}
We apply the Monte Carlo simulation of the classical perceptron with Ising weights~\cite{Gardner:unfinished} to the quantum perceptron. Here, we elaborate the details of the Monte Carlo simulation.

The first pattern $i^1_j = \pm 1$ is chosen at random and we fix a certain threshold $\kappa$. Then, we go through all possible realizations of the weights and keep only the weights that satisfy the given threshold $\kappa$. This forms the remaining set of the weights. Then, a second pattern is chosen at random and we go through all possible realizations of the remaining set of the weights to keep again only the subset of weights which satisfy the given threshold $\kappa$. Then, we continue by choosing more random patterns and updating the set of the weights that fulfill the given threshold. After a certain number of $P$ patterns that have been introduced to the perceptron, no choice for the weights exist for $P+1$ patterns. This means that for this sample the system can store exactly $P$ patterns. 

Therefore, the value of $P$ depends on the random choices of the $\sum i_{j}^{\mu} = \pm 1 $ and the threshold $\kappa$. The threshold $\kappa$ in the classical case is zero, since the output is either $\pm 1$. For the quantum case, the threshold $\kappa$ is $m/2$ since $\kappa \in [0, m]$. We need to average $P$ over many samples and define the estimate of the storage capacity for a system of size $N_s$
\begin{equation}
\alpha (N_s) = \frac{\braket{P}}{N_s}.
\end{equation}
For the numerical simulations in Fig.~\ref{MC}, $N_s$ is equal to $N$ and $m$ for the classical and quantum percepton, respectively. Moreover, we have $2^N$ and $2^m$ realizations of the weights for the classical and quantum models, respectively. We used $10000$ samples for each simulation and we performed them three times to estimate the error. In Fig.~\ref{MC}, we choose odd values of $N$ to always have $\pm 1$ for the output and we use $m={8, 16, 32} $ since for larger $m$ the computation becomes intractable.

\section{Speculations about the Gaussian inputs}\label{appendix_D}
In the derivation of the basic expression 
\begin{multline}
g= \alpha \left[ \dfrac{ \kappa}{2} - \dfrac{1}{2}  q^2 ( 2+ \kappa)^2  \right] + \dfrac{1}{2} \log \left( 1-q\right) + \dfrac{1}{2 \left( 1-q\right)},
\end{multline}
where we used an expansion in $q$ to study 
the $q\rightarrow 1$ behaviour.
The next order contribution in the effective potential is presumably
$$ \left[ \dfrac{ \kappa}{2} - \dfrac{1}{2}  q^2 ( 2+ \kappa)^2  \right] \approx \left[\dfrac{ \kappa}{2}+ \dfrac{1}{2} ( 2+ \kappa)^2 - \dfrac{1}{2 (1-q^2)} ( 2+ \kappa)^2\right]$$
and suggests that 
$$
\alpha_c( 2+ \kappa)^2/2 = 1, 
$$
implying maximal $\alpha_c(\kappa=0)= 1/8$.

\newpage
\bibliography{biblio}

\end{document}